\documentclass[a4paper]{jpconf}

\usepackage{graphicx}
\usepackage[caption=false]{subfig}
\usepackage{bm}
\usepackage{amsmath}
\usepackage{times, txfonts}

\newcommand{\nmax}{N_{\mbox{\footnotesize{max}}}}

\begin{document}
\title{Recent Applications of Self-Consistent Green's Function Theory to Nuclei}

\author{Carlo Barbieri, Francesco Raimondi and Christopher McIlroy}

\address{Department of Physics, University of Surrey, Guildford GU2 7XH, United Kingdom}

\ead{C.Barbieri@surrey.ac.uk}

\begin{abstract}
We discuss recent \emph{ab initio} calculations based on self-consistent Green's function theory.
It is found that a simple extension of the formalism to account for two-nucleon scattering outside the model space allows to
calculate non-soft interactions. With this, it is possible to make predictions for Lattice QCD potentials, obtained so far
at pion masses of $m_\pi = 0.47$~GeV/c$^2$. More traditional calculations that use saturating chiral EFT forces yield
a good description of nuclear responses and nucleon knockout spectroscopy. 
\end{abstract}

\emph{Ab initio} nuclear theory has seen remarkable advances in the last 15 years. These resulted from the combination of improved theories of realistic two-nucleon (NN) and three-nucleon forces (3NFs) and of advances in many-body  computations that have reached nuclear masses of the order of~A$\sim$100~\cite{Hebeler2015ChiralRev}.

The self consistent Green's function (SCGF) approach is based on the one-body propagator, $g(\omega)$, that describes the evolution of particle and hole states in (i.e., nucleon attached or removed to/from) the exact ground state~\cite{Dickhoff2004ppnp}.
The propagator is the solution of the Dyson equation:
\begin{align}
\label{eq:Dyson}
  g_{\alpha \beta}(\omega) = g^{(0)}_{\alpha \beta}(\omega) +  \sum_{\gamma, \, \delta} \, g^{(0)}_{\alpha \gamma}(\omega) 
 \left[  \Sigma^{(\infty)}_{\gamma \delta} +   \tilde\Sigma_{\gamma \delta}(\omega) \right]
   g_{\delta \beta}(\omega) \; ,
\end{align}
where the indices $\alpha, \beta, \ldots$ label the states of a single-particle model space basis and $g^{(0)}(\omega)$ is the uncorrelated propagator, which correspond to a mean field (MF) reference state. The central quantity is the irreducible self-energy,
\hbox{$\Sigma^\star(\omega)\equiv \Sigma^{(\infty)}  +   \tilde\Sigma(\omega)$},
 which acts as an optical potential for nucleons inside the correlated medium.  One separates an enegy-independent term $\Sigma^{(\infty)}$ that describes the average MF potential felt by all particles. The dynamic part $\tilde\Sigma(\omega)$ contains the effect of correlations beyond MF. We calculate this in the third-order algebraic diagrammatic construction [ADC(3)] approximation, which is a \emph{non perturbative} resummation of $2p1h$ and $2h1p$ configurations~\cite{Barbieri2017LNP}. We generally construct the self-energy based on a reference MF state, which is derived to   best approximate $g(\omega)$ and, therefore, solve Eq.~\eqref{eq:Dyson} iteratively. The interested reader is referred to Refs.~\cite{Barbieri2017LNP,Barbieri2009Ni56prc,Soma2014GkvII} for details of the formalism and of our implementation.

Once the propagator is known, it is easy to extract information on the ground state energy, nucleon-nucleus scattering, single-particle spectroscopy, response to external probes, and so on~\cite{Barbieri2017LNP}. In the following we describe a few very recent applications based on nuclear forces obtained either directly from Lattice QCD calculations or from chiral effective field theory (EFT).

\section{Taming the hard-core of forces from Lattice QCD}

The HAL~QCD collaboration has devised an approach to extract $n$-nucleon forces from QCD simulation of $3n$ quarks in a space-time lattice~\cite{Ishii:2006ec,Aoki:2012tk,Doi:2011gq}.  This method generates systematically consistent two-, three- and many-nucleon interactions that are faithful to the few-body data and scattering phase shifts by construction.
An earlier set of potentials was derived in the flavour SU(3) limit with different masses of the (pion) pseudo-scalar meson. Among these the force with lightest value of $M_{PS}=$~469~MeV/c$^2$ shows saturation on nuclear matter~\cite{Inoue:2013nfe}.  We refer to this as the HAL469$_{SU(3)}$ interaction.
Note that more advanced interactions at near the physical pion mass are currently being computed~\cite{Doi:2015oha}.

For forces with a strong short-range repulsion, like the HAL~QCD interactions, usual truncations of the oscillator space (of up to 12 shells in this case) are not sufficient and a resummation of ladder diagrams outside the model space is required. We do this by solving the Bethe-Goldstone Equation (BGE) in the excluded space according to Refs.~\cite{HJORTHJENSEN1995125,engeland2008cens} and add the corresponding diagrams to the mean field (MF) term of the self-energy, which becomes energy dependent~\cite{Barbieri2009Ni56prc}: 
\begin{align}
\Sigma_{\alpha\beta}^{(\infty)}(\omega) ={}& \sum_{\gamma \, \delta}\int\frac{d\omega'}{2\pi i}T^{BGE}_{\alpha\gamma, \, \beta\delta}(\omega+\omega') \, g_{\delta \gamma}(\omega') \, e^{i \omega' \eta} 
\label{eq:SigMF}
\end{align}  
where $T^{BGE}_{\alpha\gamma, \, \beta\delta}(\omega)$ are the elements of the scattering t-matrix in the excluded space. A static effective interaction is then extracted that we use to calculate the ADC(3) self-energy within the \hbox{included} model space.
To do this, we  solve the Hartree-Fock (HF) equations with the MF potential of Eq.~\eqref{eq:SigMF}:
\begin{align}
\sum_\beta  \left\{ 
  \langle\alpha|\frac{p^2}{2 m}|\beta\rangle +  \Sigma_{\alpha\beta}^{(\infty)}(\omega = \varepsilon_{r}^{HF})
 \right\} \; \psi^r_\beta 
 ={}&   \varepsilon_{r}^{HF}  \, \psi^r_\alpha \; ,
\label{eq:GHF}
\end{align} 
where latin indices label HF states. We then define a static interaction in this HF basis similarly to Refs.~\cite{Barbieri2009Ni56prc,PhysRevC.66.044301}:
\begin{align}
V_{r s, p q} ={}& \frac{1}{2}\left[ T^{BGE}_{r s, p q}(\varepsilon_{r}^{HF} + \varepsilon_{s}^{HF}) + T^{BGE}_{r s, p q}(\varepsilon_{p}^{HF} + \varepsilon_{q}^{HF})  \right] \; .
\label{eq:Veff}
\end{align}
It should be noted that the BGE used to generate $T^{BGE}(\omega)$ resums scattering states where at least one nucleon is outside the whole model space, while full ADC(3) correlations are computed for all nucleons inside the space. Hence, $T^{BGE}(\omega)$  does not suffer from ambiguities with the choice of the single-particle spectrum at the Fermi surface encountered with the usual G-matrix used in Brueckner HF calculations.
%

\begin{figure}[t]
\begin{center}
\includegraphics[height=0.3\columnwidth,clip=true]{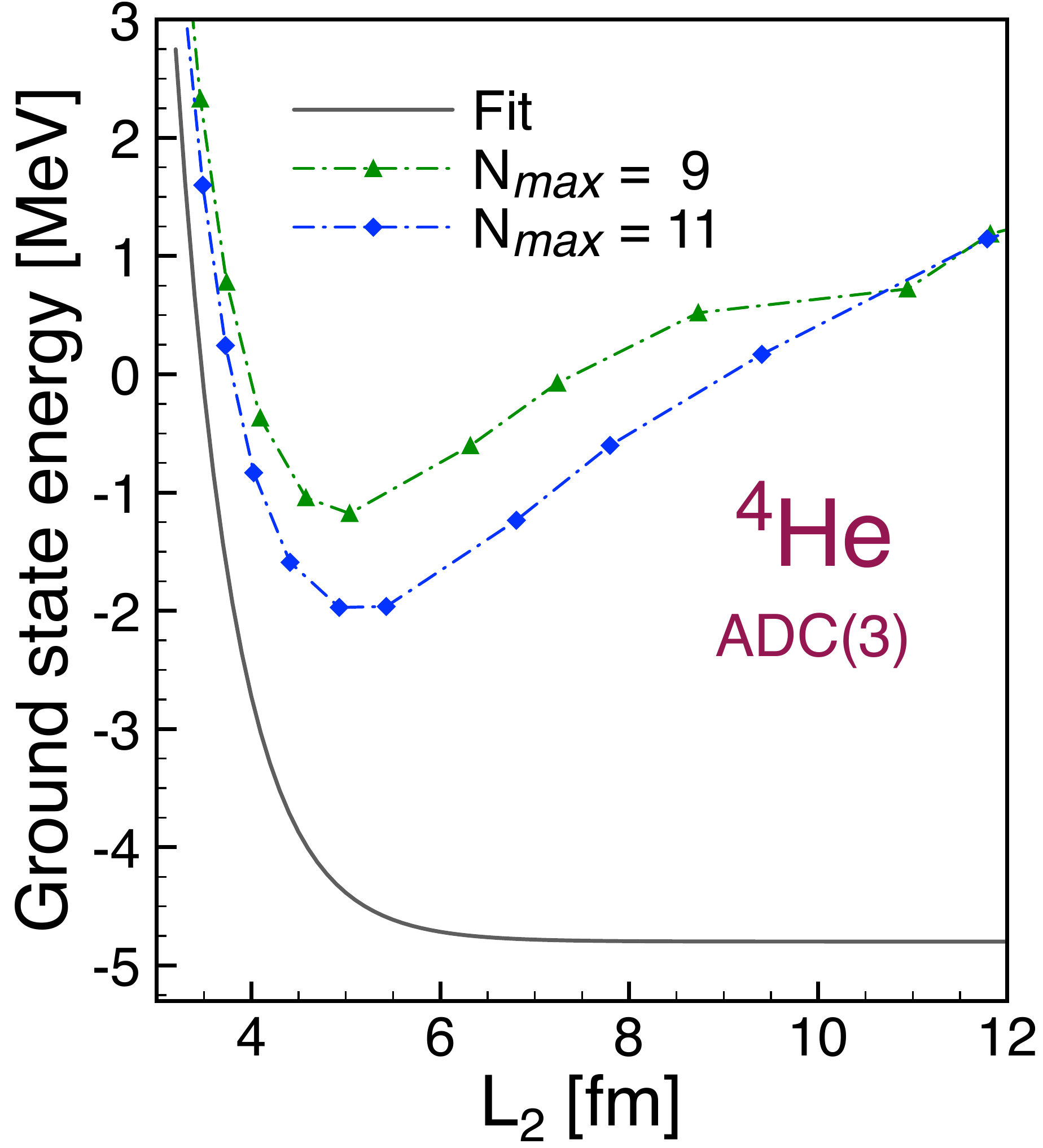} \hspace{2 cm}
\includegraphics[height=0.3\columnwidth,clip=true]{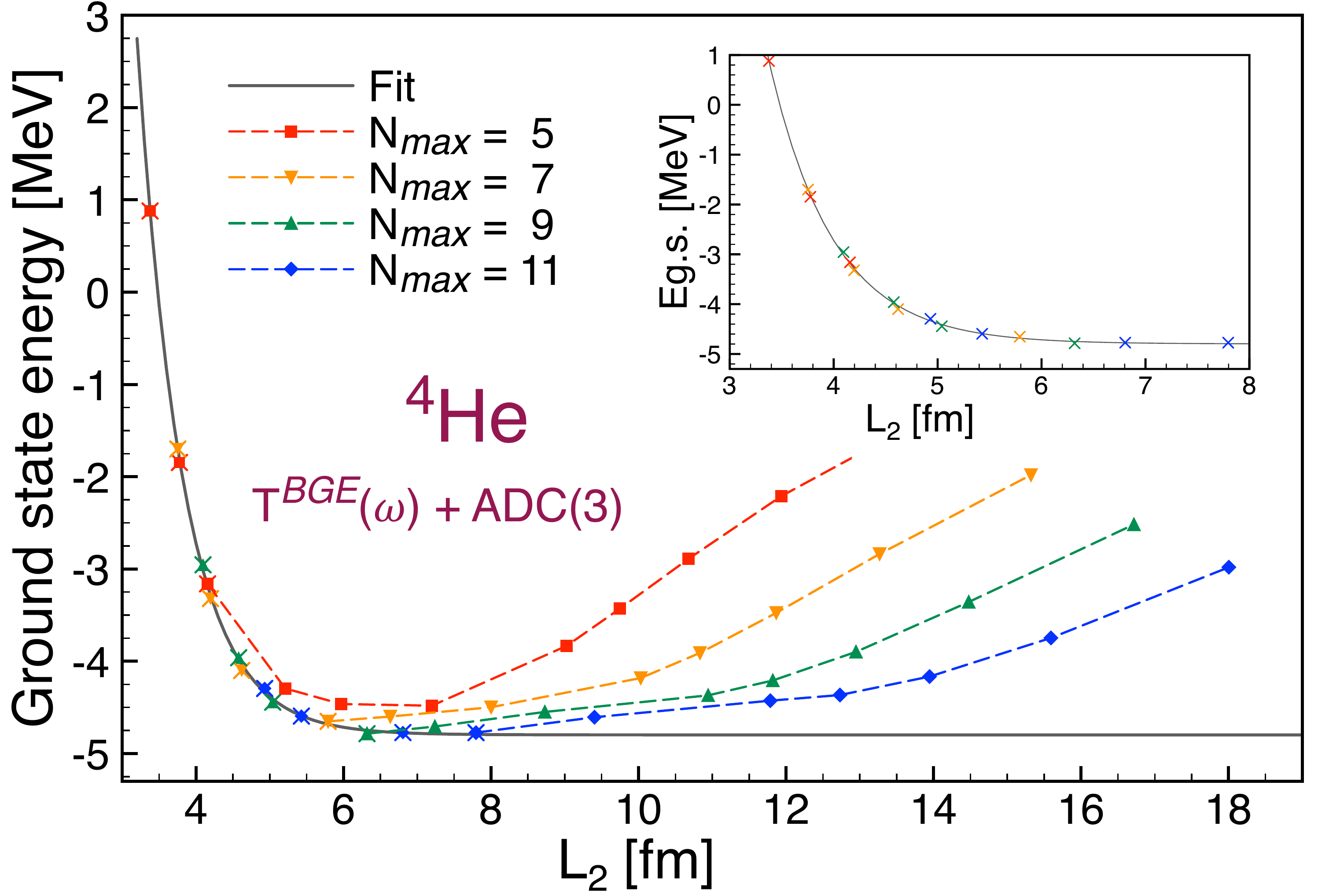}
\caption{\small 
  Calculated ground state energies of $^4$He for the HAL469 potential as a function of the effective radius $L_2$. 
 \emph{Left}:~Solution for the bare interaction at $\nmax$=9 and 11 and varying oscillator frequencies without ladders from the excluded space. 
 \emph{Right}:~Full calculation, including all ladder diagrams from outside the model space. Different colors and broken lines are a guide to the eye
  connecting results of the same  $\nmax$.  The data points included in the fit are marked with crosses and are also shown separately in the inset. For \emph{both panels}, the full black line is the result of the IR extrapolation, with the inclusion of $T^{BGE}$ ladders.}
\label{fig:H4extrap}
\end{center}
\vskip -.7 cm
\end{figure}

The combination of the ladder diagrams contained in the ADC(3) expansion and those in the $T^{BGE}(\omega)$ accounts for the complete diagonalization of short-distance degrees of freedom.  One can then study the infrared (IR) convergence of total binding energies to a complete basis set. Ref.~\cite{More2013InfraredExtr} established that a harmonic oscillator (HO) model space, of frequency $\hbar\Omega$ and truncated to the first $\nmax+1$ shells, behaves as a hard wall spherical box of radius
\begin{align}
  L_2 ={}& \sqrt{2 ( \nmax + 3/2 + 2)} \, b \; ,
\end{align}
where $b\equiv\sqrt{\hbar c^2/ m_N \Omega}$ is the oscillator length ($\hbar$=$c$=1). 
 Given an interaction that is independent of the model space, if the ultraviolet (UV) degrees of freedom are diagonalised exactly then the calculated ground state energies are expected to converge exponentially when increasing the effective radius $L_2$:
\begin{align}
  E_0^A[\nmax,\hbar\Omega] ={}& E_{\infty} ~+ ~C \; e^{- 2 \, k_\infty \,  L_2}.
\label{eq:IRextr}
\end{align}
For the bare HAL469 interaction, if we use the SCGF without ladder diagrams from outside the model space, the extrapolation according to Eq.~\eqref{eq:IRextr} will fail because the short-distance repulsion requires extremely large model spaces ($\nmax>>$ 20) to reach UV convergence, see the left panel of Figure~\ref{fig:H4extrap}. The results with $T^{BGE}(\omega)$ and the interaction~\eqref{eq:Veff} included are displayed in the right panel and show good IR convergence. The total energy is extracted  from a nonlinear least-squares fit to Eq.~\eqref{eq:IRextr}. 


\begin{figure}[t]
\begin{center}
\includegraphics[height=0.23\columnwidth,width=0.32\columnwidth,clip=true]{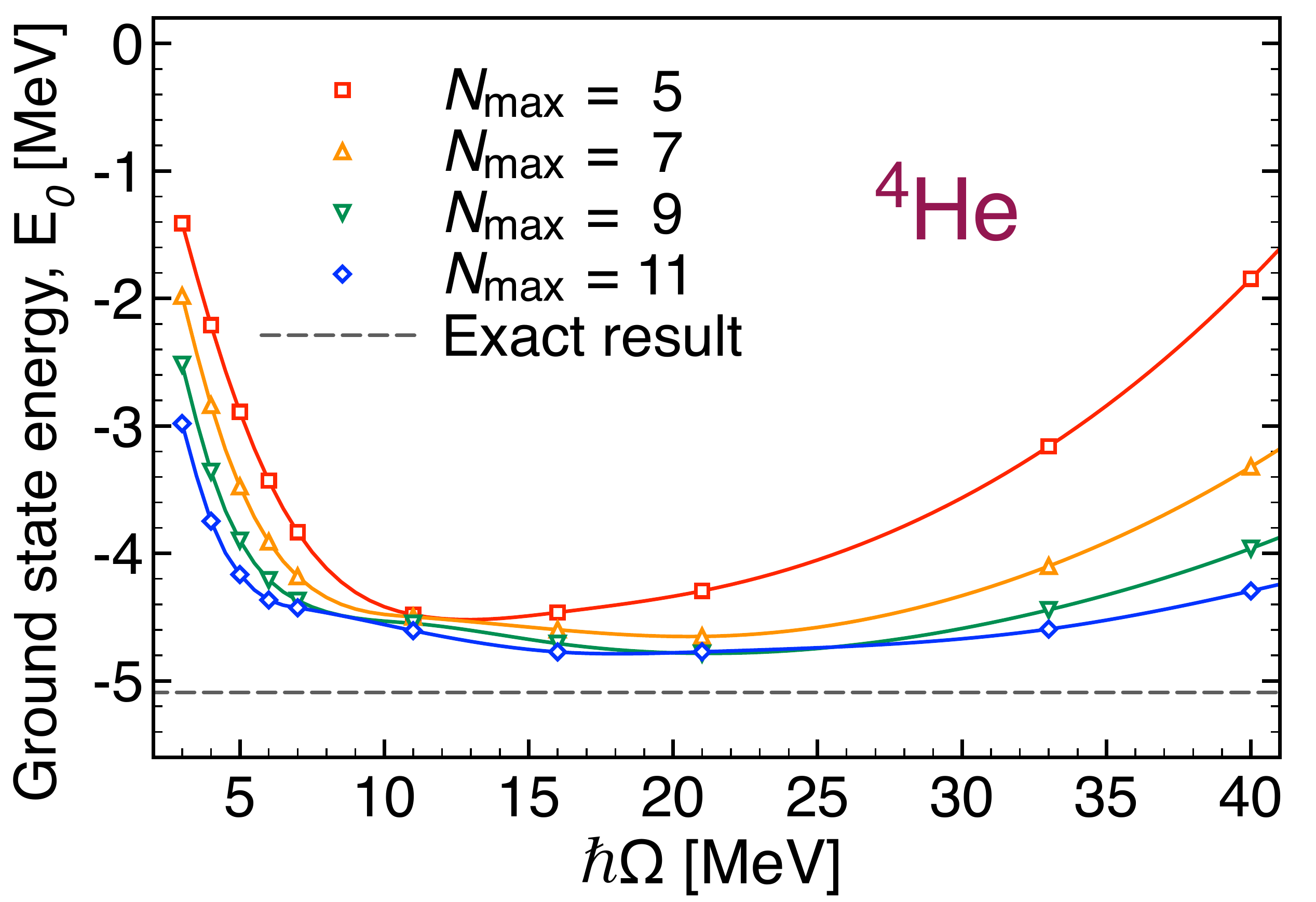} 
\includegraphics[height=0.23\columnwidth,width=0.32\columnwidth,clip=true]{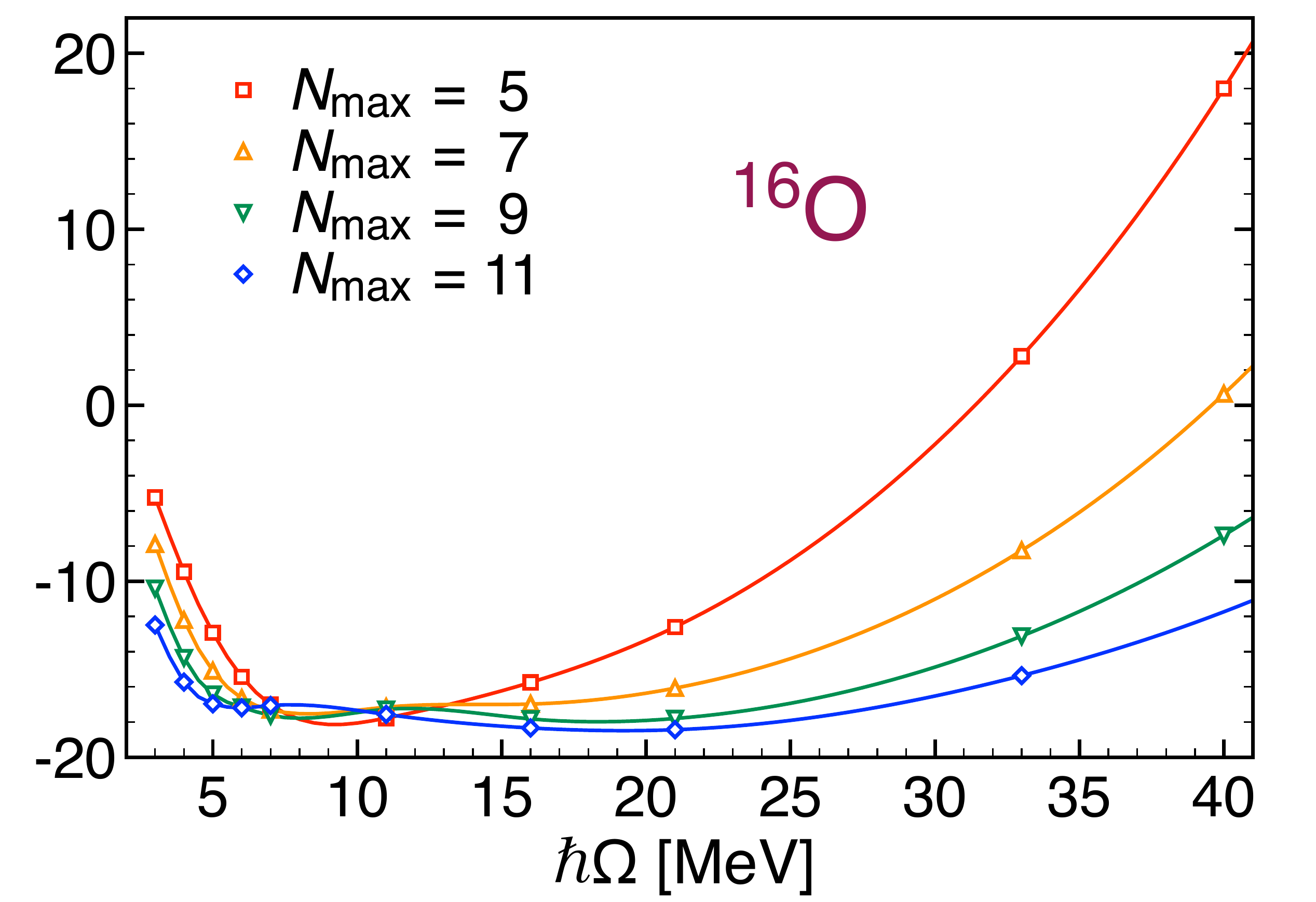} 
\includegraphics[height=0.23\columnwidth,width=0.32\columnwidth,clip=true]{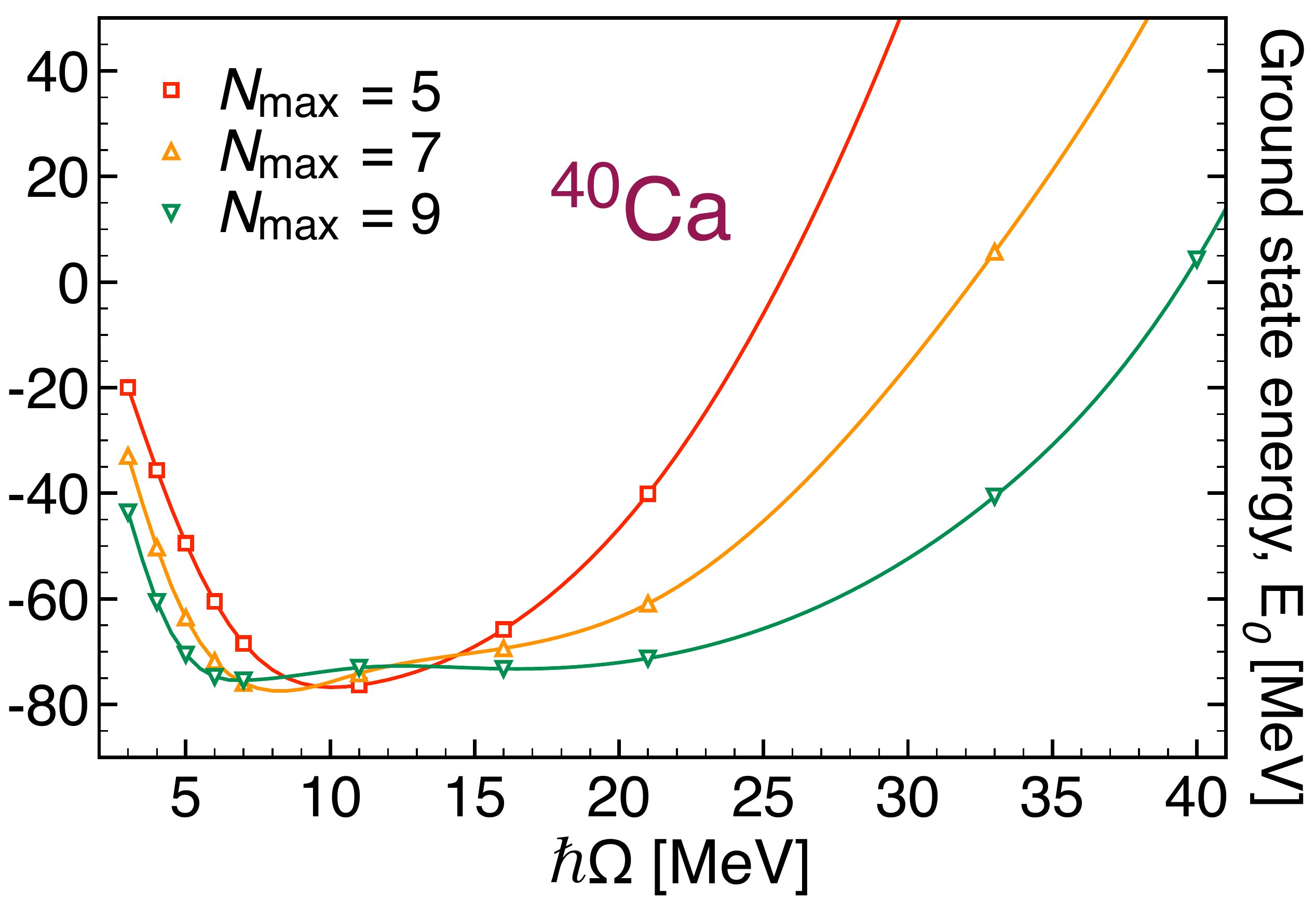}
\caption{\small 
Ground state energy of $^4$He, $^{16}$O and $^{40}$Ca as a function of the harmonic oscillator frequency, $\hbar\Omega$, and the model space size, $\nmax$.
All results are for the HAL469 potential from full self-consistent calculations in the $T^{BGE}(\omega)$ plus ADC(3) approach.
}
\label{fig:He4O16Ca40}
\end{center}
\vskip -.7 cm
\end{figure}

The one-body propagators of $^4$He, $^{16}$O and $^{40}$Ca have been calculated using $T^{BGE}(\omega)$ to remove UV modes. We employed  spherical harmonic oscillator spaces of different frequencies, $\hbar\Omega$, and sizes up to \hbox{$\nmax$=$\max\{2 n + \ell\}=$11} (and $\nmax\leq$ 9 for $^{40}$Ca) and we used the Koltun sum rule to extract the intrinsic ground state energies from $g(\omega)$.
Fig.~\ref{fig:He4O16Ca40}  shows that the complete resummation of ladder diagrams outside the model space results in a somewhat flat behaviour of the total energies for \hbox{$\hbar\Omega\approx$5-20~MeV.}  While some oscillations w.r.t. $\hbar\Omega$ are still present, the IR convergence pattern (shown in Fig.~\ref{fig:H4extrap} for $^4$He) is very clean for all three isotopes.
This gives confidence that the short-range repulsion of HAL469 is relatively mild and that it is accounted for accurately.
 From calculations up to $\hbar\Omega$=50~MeV and the IR extrapolation, we estimate a converged binding energy of 4.80(3) MeV for $^4$He, where the error corresponds to the uncertainties in the extrapolation.  This is to be benchmarked on the \emph{exact} result of -5.09~MeV, which is known from Stochastic Variational calculations~\cite{Nemura2016WSPC_procs}. Since the method is size extensive, we assume  a 10\% error due to many-body truncations for all isotopes, as a conservative estimate. 
 
We obtain -17.9(0.3)(1.8)~MeV for the ground state energy of $^{16}$O, where the first error is from the IR extrapolation. Hence, $^{16}$O is unstable with respect to 4-$\alpha$ break up, by~$\approx$2.5~MeV, although allowing an error in our binding energies of more than 10\% could make it slightly bound. This is in contrast to the experiment, at the physical quark masses, where the 4-$\alpha$ breakup requires 14.4~MeV. On the other hand, $^{40}$Ca we calculate -75.4(6.7)(7.5)~MeV and this is stable with respect to $\alpha$ breakup by $\approx$24~MeV.

\section{Isovector Dipole Nuclear Response}

In this Section we focus on the $^{16}$O, $^{22}$O, $^{40}$Ca and $^{48}$Ca nuclear response produced by an isovector dipole electric field, E1.  The corresponding operator, corrected for the center-of-mass displacement, is:
\begin{equation}
\label{F1}
\hat{\mathcal{Q}}_{1m}^{T=1}=\frac{N}{N+Z} \sum_{p=1}^{Z} r_p Y_{1m} - \frac{Z}{N+Z} \sum_{n=1}^{N} r_n Y_{1m} \, ,
\end{equation}
which probes the excitation spectrum with multipolarity and  parity $J^{\pi}$=1$^{-}$. 
We calculate the response function $R(E)$ of Eq.~\eqref{F1} by solving the usual random phase approximation (RPA) in the particle-hole channel and starting from the same MF propagator that is used as reference state to obtain the self-energy in previous sections. As mentioned in the introduction, this includes effects of correlations that go beyond the usual HF mean field and, in  particular, those responsible for reproducing the correct centroid of giant resonances. 
%
%
We then compute the total photoabsorption 
cross section as
\begin{equation}
\label{CrossSec}
\sigma(E)= 4\pi^2 \alpha \,  E \, R(E) \, 
\end{equation} 
and the dipole polarizability, which is the  accumulated E1 strength weighted by the inverse of the energy:
\begin{equation}
\label{polariz}
 \alpha_{\text{D}}= 2 \alpha  \int_{E_{thr}}^{+\infty} dE \, \frac{R(E)}{E} \, .
\end{equation} 
In the above relations,  $\alpha$ denotes the fine-structure constant.


\begin{figure}[t]
  \centering
    \subfloat[]{\label{O16}\includegraphics[scale=0.35]{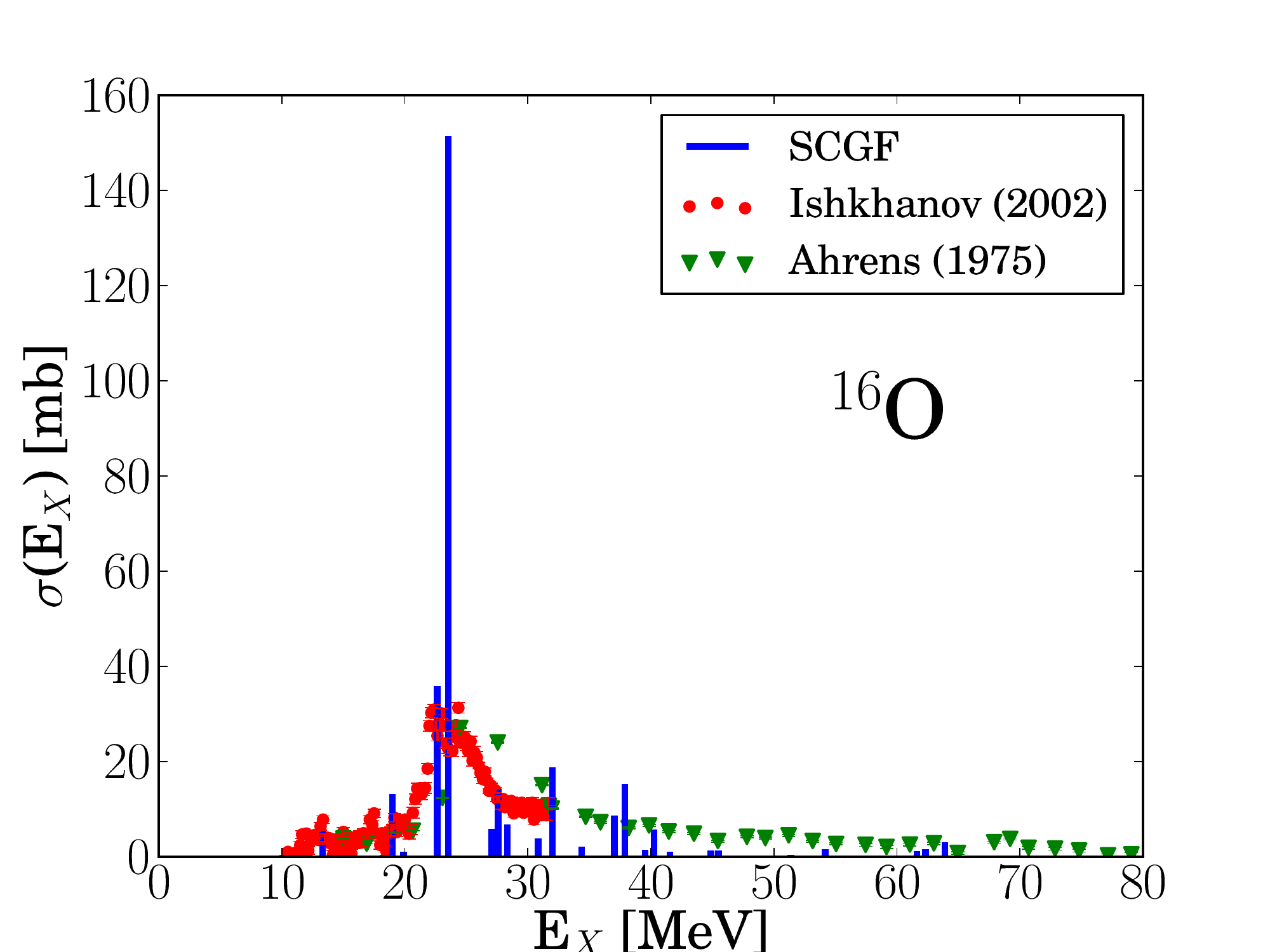}}
   \subfloat[]{\label{O22}\includegraphics[scale=0.35]{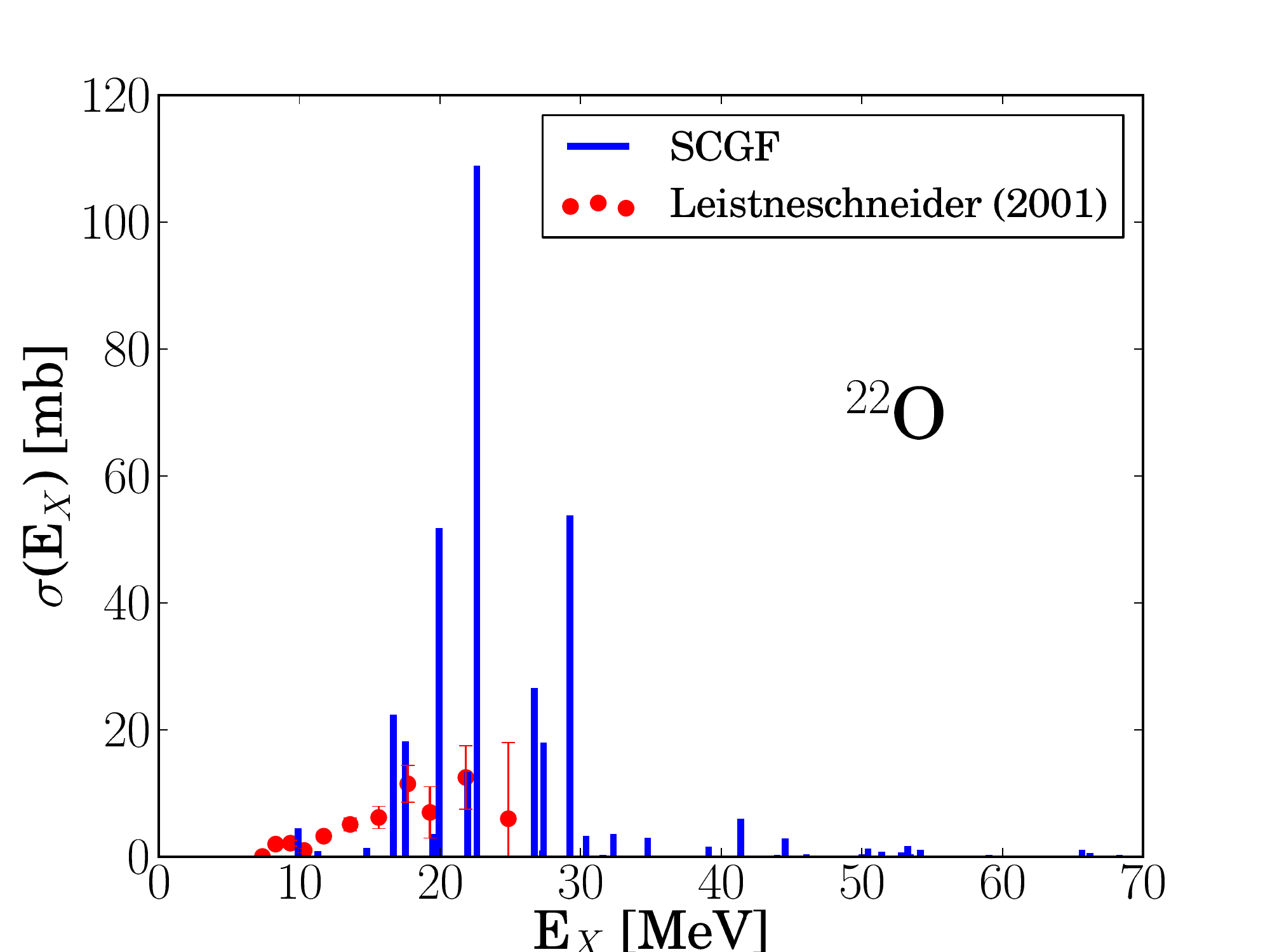}}
  \caption{\label{O16O22}\small Computed total cross section for $^{16}$O (a) and $^{22}$O (b) compared to experimental data obtained from photoabsorption~\cite{Ishkhanov,AHRENS1975479} and Coulomb excitation~\cite{PhysRevLett.86.5442} experiments, respectively. The size of the HO basis in the calculation is $\nmax$ = 13 (i.e., 14 major shells) and frequency $\hbar\Omega$ = 20 MeV. The interaction used is NNLO$_{\text{sat}}$.}
  \label{Oxy}
\vskip -.5 cm
\end{figure}

\begin{figure}[t]
  \centering
    \subfloat[]{\label{Ca40}\includegraphics[scale=0.35]{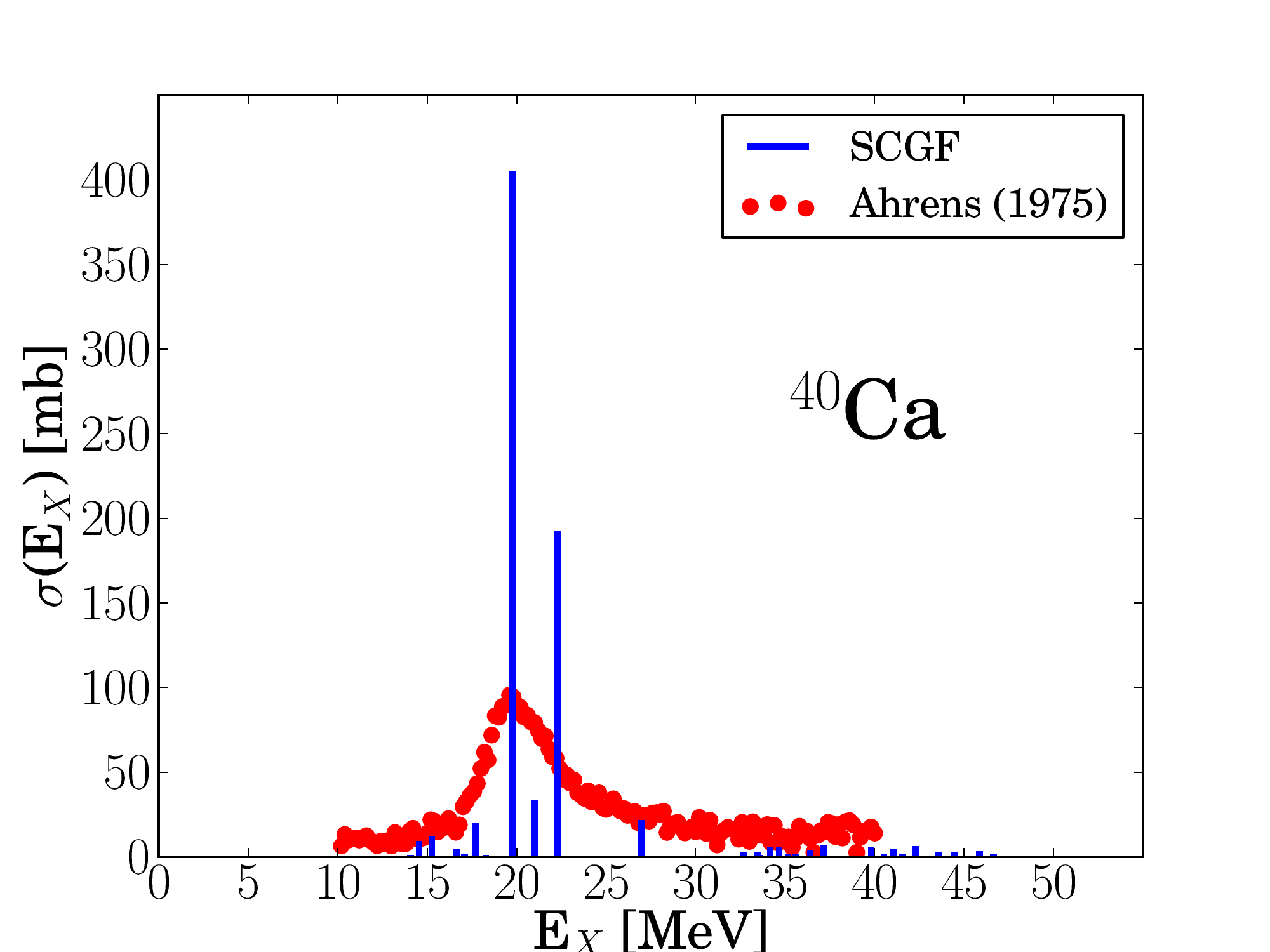}}
   \subfloat[]{\label{Ca48}\includegraphics[scale=0.35]{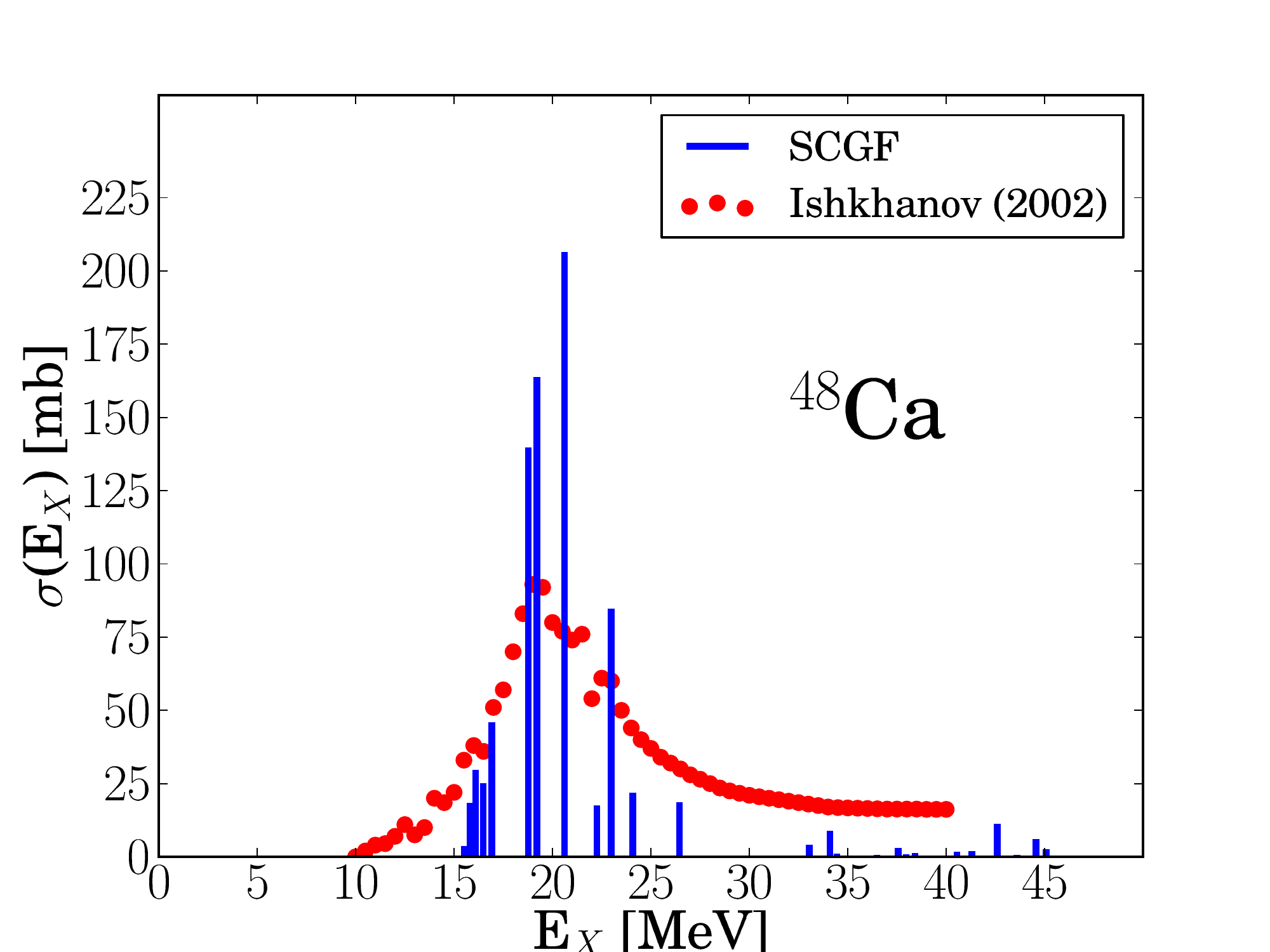}}
  \caption{\label{Ca40Ca48}\small Same as Fig.~\ref{Oxy} but for $^{40}$Ca (a) and $^{48}$Ca (b). Experimental data are taken from Ahrens \textit{et al}~\cite{AHRENS1975479} and Ishkhanov \textit{et al}~\cite{Ishkhanov} respectively. }
  \label{Calcium}
\vskip -.5 cm
\end{figure}

\begin{table}[b]
\begin{center}
\begin{minipage}{17pc}
\begin{tabular}{ |c|c|c|c| } 
\hline
Nucleus & SCGF & CC-LIT & Exp \\
\hline
$^{16}$O & 0.50 & 0.57(1) & 0.585(9) \\ 
$^{22}$O & 0.72 & 0.86(4) & 0.43(4) \\ 
$^{40}$Ca & 1.79 & 1.87 & 1.87(3) \\
$^{48}$Ca & 2.08 & 2.45 & 2.07(22) \\
\hline
\end{tabular}
\end{minipage}
\begin{minipage}{20pc}
\caption{\small Isovector dipole polarizabilities $\alpha_{\text{D}}$ for $^{16}$O, $^{22}$O, $^{40}$Ca and $^{48}$Ca, computed in the SCGF approach (second column), the CC-LIT approach (third column) and obtained by integrating the experimental data (fourth column). \label{TabPol}
}
\end{minipage}
\end{center}
\end{table}

Our results for the total cross sections for $^{16}$O, $^{22}$O, $^{40}$Ca and $^{48}$Ca are compared with  photoabsorption and Coulomb excitation experiments in Figs.~\ref{O16O22} and~\ref{Ca40Ca48}. The computed spectrum is collected into peaks since we diagonalize the RPA matrix in a finite model space.  We performed calculations using the NNLO$_{\text{sat}}$ chiral interaction that reproduces radii in this region and has good saturation properties~\cite{Ekstrom2015NNLOsat}.  For all isotopes considered, the position of the giant dipole resonance is fairly well reproduced.

By moving from the N=Z nucleus $^{16}$O  to the neutron-rich $^{22}$O, the strength of the response is distributed  towards lower energies, with the first RPA peak shifted to energy below 10 MeV. In the corresponding experimental curve, there appears a soft dipole mode of excitation, with a weaker strength compared to the giant resonance at higher energies. 

In Table~\ref{TabPol},  dipole polarizabilities $\alpha_{\text{D}}$ of Eq.~(\ref{polariz}) are compared  with values obtained from the  integrated experimental spectrum and from a Coupled Cluster combined with the Lorentz Integral Transform (CC-LIT) approach calculation~\cite{PhysRevC.94.034317,PhysRevLett.118.252501}. 
For the  $^{16}$O, where the experimental dipole polarizability is fully integrated over the entire spectrum, our computed $\alpha_{\text{D}}$ lacks 15\% of the strength with respect to both the experimental value and result of the CC-LIT approach. The total strength in the SCGF calculation is better recovered for higher mass $^{40}$Ca and $^{48}$Ca, and even within the experimental error for~$^{48}$Ca.
Correcting the discrepancies with CC-LIT may require improvements of the many-body truncation to go beyond the RPA and we are currently working to implement such extensions. 

\section{Spectroscopic factors}

\begin{figure}
\begin{center}
\includegraphics[width=.4\columnwidth,clip=true]{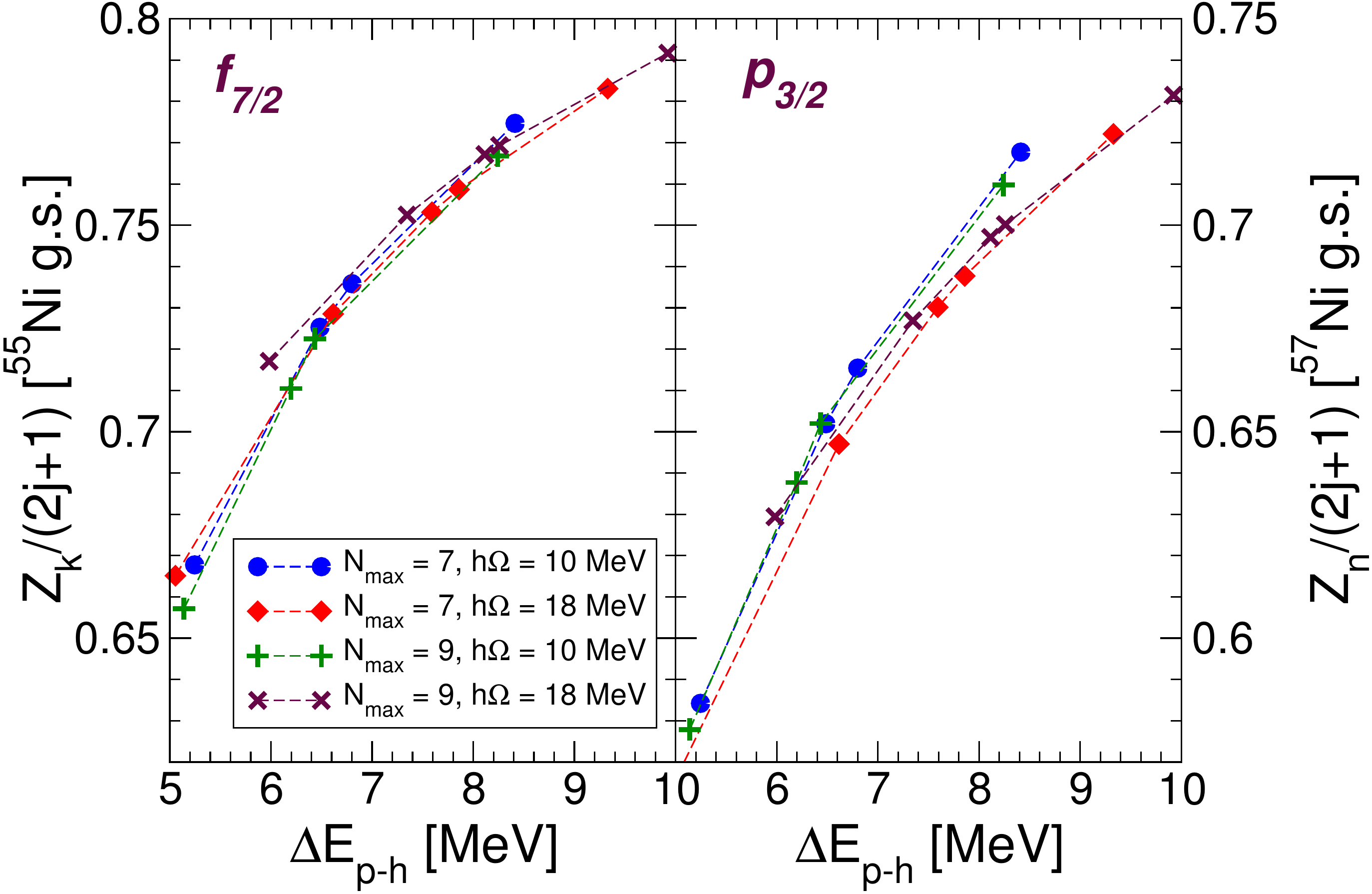}  \hspace{1cm}
\includegraphics[width=.38\columnwidth,clip=true]{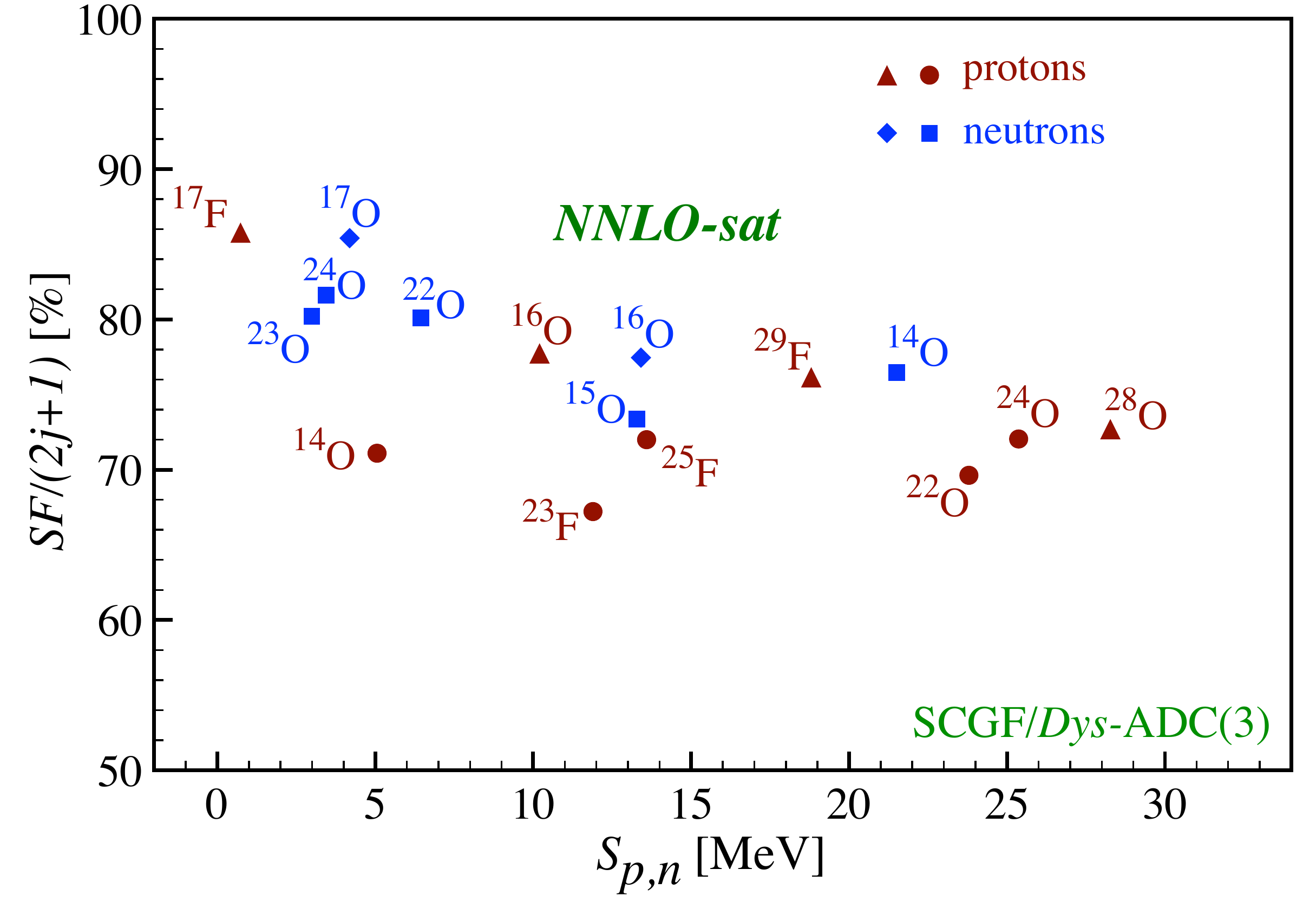}    
\caption{\small {\em Left.} Dependence of spectroscopic factors of the first quasiparticle and quasihole peaks in $^{56}$Ni as a function of the particle-hole gap $\Delta E_{ph}$~\cite{Barbieri2009Ni56prc}. The calculated quenching is strongly correlated to the gap even when full convergence with respect to the model space is still not achieved. Note that $\Delta E_{ph}\equiv E^{A+1}_0-E^{A-1}_0$ is an experimentally observable quantity. Here, it is varied over a range of values by tuning the monopoles of the NN interaction. 
  {\em Right.}  Calculated spectroscopic factors for protons and neutrons around Oxygen isotopes obtained from NNLOsat.  Each point refer to the separation of a nucleon from the isotope indicated nearby to the ground state of the daughter nucleus. Likewise, stronger quenching is found for those isotopes with smaller~$\Delta E_{ph}$ gap. }
\label{Fig:SFs_vs_Eph}
\end{center}
\vskip -0.7 cm
\end{figure}

  The fragmentation of the single-particle spectral function is mostly determined by long-range correlations (LRC) and it is constrained by general properties at the energy surface, such as the density of states and gaps~\cite{Barbieri2009PRL}.  As an example, Fig.~\ref{Fig:SFs_vs_Eph} demonstrates the dependence of spectroscopic factors (SFs) for the dominant quasiparticle peaks in $^{\rm 56}$Ni  on the  particle-hole gap $\Delta E_{ph}$ (which has an experimental value of 6.1~MeV).  This is calculated from a chiral N3LO two-nucleon interaction only but with a modification of its monopole strength in the $pf$ shell.  The monopole correction controls the separation among the $p_{3/2}$ and the  $f_{7/2}$ orbits. Hence, by varying its strength one can control the predictions for $\Delta E_{ph}$. Calculations for spaces of different sizes and HO frequencies are not converged w.r.t. the model space but all lines lay on top of each other, showing that there exists a clear correlation and that SFs can be strongly constrained by observable quantities at low energy. 
   
   Since the correlation of SFs with $\Delta E_{ph}$ is strong, we calculated the oxygen chain using  NNLOsat Hamiltonian of Ref.~\cite{Ekstrom2015NNLOsat}, which has the advantage of predicting gaps accurately. The spectroscopic factors obtained in the ADC(3) scheme are shown in the right panel of Fig.~\ref{Fig:SFs_vs_Eph} and are sensibly smaller than previous results from older Hamiltonians that had a too dilute spectrum at the Fermi surface~\cite{Cipollone2015OxSpFnct}. Remarkably, the NNLOsat results are very close (almost equal) to past FRPA calculations of $^{\rm 16}$O where the quasiparticle energies were phenomenologically constrained to their experimental values~\cite{Barbieri2002o16}.
  These are also in good agreement with ($p$,$2p$) measurements from R$^3$B at GSI~\cite{AumannIschia}. 
 
\ack
{\small We thank the HAL QCD collaboration for providing the HAL469$_{\hbox{\small  \em SU(3)}}$ interaction.  
This work was supported in part by the United Kingdom Science and Technology Facilities Council (STFC) under Grants No. ST/L005743/1 and ST/L005816/1 and by the Royal Society International Exchanges Grant No.~IE150305. Calculations were performed at the DiRAC Complexity system at the University of Leicester (BIS National E- infrastructure capital Grant No. ST/K000373/1 and STFC Grant No. ST/K0003259/1).}

\section*{References} 
 \bibliography{Procs_Ischia}

\end{document}